\documentclass[11pt,twocolumn]{article}

\usepackage[utf8]{inputenc}
\usepackage[spanish,english]{babel}
\usepackage{amsmath,amssymb}
\usepackage{authblk}
\usepackage{geometry}
\usepackage{abstract}
\usepackage{natbib}
\usepackage{hyperref}
\usepackage{fancyhdr}
\usepackage{parskip}

\usepackage{etoolbox} 
\AtBeginEnvironment{onecolabstract}{\setlength{\parskip}{12pt}} 

\hypersetup{
    colorlinks=true,
    linkcolor=blue,
    filecolor=magenta,      
    urlcolor=cyan,
    pdftitle={Título del artículo},
    bookmarks=true
}

\title{\textbf{Consideraciones para una formulación termodinámica neo-gibbsiana aplicada a sistemas sociales}}
\author[1]{Guzmán R. Victor A.\thanks{Correo: vguzman4@uc.edu.ve}}
\author{Solano H. Esneiber H.}
\affil[1]{Departamento de Física, Facultad de Ciencias y Tecnología, Universidad de Carabobo. Venezuela}
\date{} 

\begin{document}

\twocolumn[
  \maketitle
  \vspace{-24pt}
  \renewcommand{\abstractname}{Resumen}
  \begin{onecolabstract}
  \selectlanguage{spanish}
  Este trabajo propone un primer acercamiento hacia una teoría termodinámica neo-gibbsiana para sistemas sociales, basada en parámetros económicos y culturales cuantificables, evitando el uso de analogías directas con las ecuaciones de estado de la termodinámica clásica. Para ello se revisan las consideraciones establecidas por propuestas termodinámicas para modelar la economía y se establece un marco conceptual para la definición de parámetros intensivos capaces de describir equilibrios y dinámicas evolutivas en sociedades, con especial atención a la diversificación cultural y estructural de bienes, recursos y al fenómeno del multimonetarismo. La relación entre las variables se hace introduciendo como un primer acercamiento una función no decreciente $S$, definida para todos los procesos sociales, que actúa como medida de cantidad de leyes, normas y otros signos simbólicos cuantificables para toda sociedad. Finalmente, se analiza su interpretación en relación con fuerzas termodinámicas, particularmente en el contexto del equilibrio que se da para dos sistemas sociales simples que interactúan intercambiando recursos.
  
  \textbf{Palabras clave:} Termodinámica \and Sociología \and Sistemas Complejos \and Entropía Social.
  \end{onecolabstract}

  \vspace{0.5em}
  \renewcommand{\abstractname}{Abstract} 
  \begin{onecolabstract}
  \selectlanguage{english}
    This work proposes a first approach towards a neo-Gibbsian thermodynamic theory for social systems, grounded in quantifiable economic and cultural parameters, while deliberately avoiding direct analogies with classical thermodynamic equations of state. Building on previous thermodynamic approaches to economic modeling, we introduce a conceptual basis for defining intensive variables capable of characterizing equilibrium states and evolutionary dynamics in societies. Particular attention is given to cultural and structural diversification of goods and resources, as well as to the phenomenon of multi-currency systems. A non-decreasing function $S$ is defined over all social processes to represent the accumulation of laws, norms, and other quantifiable symbolic elements. Finally, we examine its interpretation in terms of thermodynamic forces, focusing on the case of equilibrium between two simple, interacting social systems exchanging resources.
  
  \end{onecolabstract}
  
  \vspace{1em}
]

\selectlanguage{spanish}

\section{Introducción}

En esta investigación se exploran los elementos necesarios para una teoría termodinámica neo-Gibbsiana para sistemas sociales en base al enfoque de \cite{callen1985} siguiendo un procedimiento análogo al seguido en \cite{Saslow1999} para sistemas económicos. El trabajo tiene una finalidad exploratoria, bajo este enfoque la idea consiste en identificar un primer conjunto de variables extensivas aditivas sobre subsistemas para sistemas sociales bajo una forma de equilibrio social, junto a alguna cantidad que sea no decreciente al evolucionar los sistemas\let\thefootnote\relax\footnote{\texttt{vguzman4@uc.edu.ve}}.

Múltiples intentos se han realizado para establecer conexiones entre la termodinámica y las ciencias sociales. En ese sentido las constribuciones de Georgescu Roegen impulsaron una ola de investigaciones entorno a esta área. En \cite{georgescu1977} se plantean múltiples elementos relevantes como la irreversibilidad en ciertos procesos sociales por medio de la idea de la degradación de la materia-energía, se entiende la entropía acá en un sentido antropomorfo al asociarla a una parte de la energía que no puede ser aprovechada socialmente.

Las características locales de las sociedades en sus inicios son significativas para entender su comportamiento a futuro como se expresa en \cite{georgescu1977}. Ya para 1978 se señala la dificultad de crear un nexo formal entre la termodinámica y la economía como ciencia social, esta investigación \cite{berry1978} explora la idea del uso de los potenciales termodinámicos para medir procesos económicos, además utiliza la idea de procesos cuasi-estáticos de \cite{callen1985} que mas adelante usa \cite{Saslow1999} para definir la evolución en el tiempo de sistemas en equilibrio.

La existencia de equivalentes termodinámicos en la economía al teorema del trabajo máximo y esta relación de la economía con la termodinámica se presentan en \cite{berry1978} no como una analogía o reinterpretación, sino de manera textual pensando en la energía y sus transformaciones por procesos industriales lo que implicaría una relación en un subconjunto de las variables económicas que describen al sistema.

La existencia de equivalentes termodinámicos en la economía al teorema del trabajo máximo y esta relación de la economía con la termodinámica se presentan en \cite{berry1978} no como una analogía o reinterpretación, sino de manera literal pensando en la energía y sus transformaciones por procesos industriales lo que implicaría una relación en un subconjunto de las variables económicas que describen al sistema.

La manera en que se plantea el nexo con la termodinámica como teoría puede llevar a que el estado de equilibrio termodinámico no coincida necesariamente con el estado de equilibrio económico, ya que en \cite{berry1978} las variables termodinámicas representan solo un subconjunto del problema económico. Otros conceptos como el de los procesos en los que una variable no cambia similar al de las isotermas, adiabáticas e isobáricas para estudiar la evolución al equilibrio de sistemas se abordan en \cite{berry1978}.

Las contribuciones en el área han sido principalmente orientadas a relacionar la termodinámica con la economía. En ese sentido se ha abordado la importancia de plantear los recursos como una variable en el modelado \cite{georgescu1979energy} y su conservación como parte del proceso. 

El uso de analogías o metáforas físicas para la descripción de sistemas económicos es señalada en \cite{mirowski1989} por ejemplo en la relación establecida entre la teoría de control óptimo y la dinámica hamiltoniana así como el uso de la suma de la utilidad mas el gasto como un invariante en analogía con la conservación de la energía, esta investigación critica la incorporación de nuevas herramientas matemáticas a la economía sin una revisión profunda de los fundamentos \cite{mirowski1989}.

El trabajo \cite{Ramirez1991} explora las funciones homogéneas Cobb-Douglas y CES así como la función de producción en función del capital y el trabajo y su interpretación en el contexto económico. La identificación de qué funciones homogéneas existen y su orden en el contexto económico y social puede usarse como indicador de candidatos a parámetros extensivos para un modelo termodinámico de la sociedad.

\section{Criterios para la selección de variables sociales}

La falta de consenso en relación a la aplicabilidad de la termodinámica para modelar los sistemas económicos-sociales ha llevado a la aparición de múltiples modelos que incluyen investigaciones rechazando las condiciones requeridas por otros modelos \cite{kovalev2016}, por parte de \cite{Binswanger1993} se señala la inaplicabilidad de la termodinámica en el equilibrio para los aspectos económicos, sugiriendo un enfoque fuera del equilibrio como el de Ilya Prigogine y señala que aunque han habido intentos fallidos no descarta la posibilidad de usar la entropía de alguna buena manera.

La intencionalidad de usar la entropía en el contexto económico es defendido en \cite{smith1996} partiendo de lo que representa física y filosóficamente. Se ha criticado el reduccionismo de buscar unificar los múltiples modelos económicos, ya que la economía como parte de la sociedad es un sistema complejo en el cual la pluralidad de perspectivas es inevitable \cite{Costanza1993}. Esta investigación señala que la idea sería buscar unos ``teoremas robustos'' como candidatos para integrar teorías.

Las ideas de fases y transiciones de fases fueron usadas en \cite{Collins1992} para estudiar sistemas económicos, en los cuales la temperatura es una medida del ritmo de actividad empresarial, acá se señala como la entropía puede ser pensada de dos manera, una es como un nivel de ruido producido por las fluctuaciones del volumen en relación a un modelo $f(t)$ que representa el ajuste por mínimos cuadrados:
\begin{equation}
    T(t):=\frac{|V(t)-f(t)|}{V(t)},
\end{equation}
la otra manera es en analogía a la ley de gases ideales $PV=NRT$, haciendo el simil de la presión $P$ con los precios:
\begin{equation}
   T(t):=\frac{PV}{N},
\end{equation}
donde $N$ representa el número de trabajadores. La temperatura es medida y distintas configuraciones estudiadas. Nótese que al $T$ no estar definida formalmente sus unidades se adaptan a la analogía que se haga en el modelo en cuestión.

Para la caracterización del estado del sistema no se usarán analogías para establecer conexiones con la termodinámica, sino que se usarán parámetros extensivos que sean aditivos bajo los subsistemas constituyentes como lo son los $X_{B,i}$ bienes, los $X_{R,i}$ recursos y las múltiples monedas que representan el dinero $X_{M,i}$. Todos estos son aspectos económicos de la sociedad, donde $x_{b}$, $x_{r}$ y $x_{m}$ representan el número de bienes, recursos y monedas. Las distintas monedas permiten pensar en sociedades bajo un sistema multimonetario.

La elección de estos parámetros es similar a la de \cite{Saslow1999} y \cite{pospelov2013} con una variación al incorporar una diversificación estructural del sistema permitiendo múltiples formas en las que pueden aparecer estas cantidades. 

Como un primer acercamiento para el modelado del aspecto social no económico se usarán las cantidades de libros $Y_{L,i}$, obras de artes plásticas $Y_{A,i}$ y canciones $Y_{M,i}$ asociadas al sistema, donde $y_{l}$, $y_{a}$ y $y_{m}$ representan la cantidad de géneros literarios, de artes plásticas y musicales respectivamente. Los parámetros $y_{l}$, $y_{a}$ y $y_{m}$ representan unas medidas de diversidad cultural que en sí mismos pueden ser identificados para cada sociedad.

Por otro lado el número $N_{i}$ de agentes o actores sociales en el sistema puede representar la cantidad de agentes o actores sociales de un grupo social. En ese sentido $n$ representa la cantidad de diferentes grupos sociales que hay, bajo la consideración de que estos grupos sociales son mutuamente excluyentes para este modelo.

Se introduce un parámetro social $S$ que toma el rol de una forma de entropía social y que representa los signos simbólicos de la sociedad en el aspecto cultural que no están representados en las formas sociales representadas previamente por los $Y_{*,i}$, estos signos simbólicos de acá son cuantificables en cantidad de normas, leyes u otras formas y cuya extensividad sobre los subsistemas no se requiere de manera general pero para esta primera aproximación se asumirá; además la misma se asume como no decreciente.

Se ha explorado previamente la naturaleza del conocimiento por medio de analogías a la termodinámica como una especie de energía o entropía \cite{bratianu2020}, en esta investigación el parámetro social $S$ jugará un rol similar al cuantificar unas formas de conocimiento. 

Finalmente en esta investigación no se explorará un enfoque mecánico estadístico basado en agentes \cite{caticha2024} o partículas para estudiar las propiedades macroscópicas, el enfoque se desarrollará utilizando parámetros extensivos macroscópicos que sirvan para caracterizar al sistema como un todo.

\section{Termodinámica social}

Se establece acá por sociedad a un sistema conformado por agentes que comparten y transforman recursos materiales y estructuras simbólicas mediante patrones normativos, institucionales y comunicativos en un espacio determinado.

El estado de un sistema social en el equilibrio se establece que está completamente caracterizado por un conjunto de parámetros extensivos económicos $\{X_{i}\}$, culturales $\{Y_{i}\}$ y por las cantidades de actores sociales $\{N_{i}\}$.

Finalmente se define una cantidad $S$ que está bien definida para todo sistema social en el equilibrio y que se puede describir en términos de los parámetros extensivos del sistema $\{X_{i}\}$, $\{Y_{i}\}$ y $\{N_{i}\}$, de manera tal que los valores asumidos por $S$ son tales que se maximizan de acuerdo a las ligaduras impuestas sobre el sistema.

Como una primera aproximación al problema se va a estudiar el caso de una sociedad sin diversidad económica, cultural o por grupos sociales, de manera que:
\begin{equation}
    S:=S(X_{R},X_{B},X_{M},Y_{L},Y_{A},Y_{M},N),
\end{equation}
donde $X_{R}$ representa la cantidad de recursos, $X_{B}$ la cantidad de bienes, $X_{M}$ la cantidad de dinero, $Y_{L}$ la cantidad de libros, $Y_{A}$ la cantidad de obras de artes plásticas, $Y_{M}$ la cantidad de música producida y $N$ la cantidad de integrantes de esa sociedad.

La evolución entre dos estados $A$ y $B$ se da de manera tal que el sistema evolucionará obedeciendo que:
\begin{equation}
    \Delta S \geq 0.
\end{equation}
Usando la idea de procesos cuasi-estáticos como en [1], [2] y [4] se puede definir $dS$ de manera que:
\begin{align}\label{dS}
    dS=&\sum_{i=1}^{3}\left(\frac{\partial S}{\partial X_{i}}\right)_{\{X_{j}\}_{j\neq i}, \{Y_{j}\}, N}dX_{i} +\nonumber \\ 
    &+\sum_{i=1}^{3}\left(\frac{\partial S}{\partial Y_{i}}\right)_{\{X_{j}\}, \{Y_{j}\}_{j\neq i}, N}dY_{i}+\nonumber\\ 
    &+\left(\frac{\partial S}{\partial N}\right)_{\{X_{j}\}, \{Y_{j}\}}dN,
\end{align}
la ecuación \ref{dS} tiene varias cosas para analizar, por un lado muestra como los cambios en los signos simbólicos $dS$ dependen de las variaciones en aspectos económicos $dX_{i}$, culturales $dY_{i}$ y poblacionales $dN$. Adicionalmente introduce unos conjuntos de parámetros intensivos que por definición son funciones homogéneas de orden cero al ser los extensivos funciones homogéneas de primer orden:
\begin{subequations}\label{Intensivos}
\begin{align}
    E_{i}&=\left(\frac{\partial S}{\partial X_{i}}\right)_{\{X_{j}\}_{j\neq i}, \{Y_{j}\}, N},\qquad\\
    C_{i}&=\left(\frac{\partial S}{\partial Y_{i}}\right)_{\{X_{j}\}, \{Y_{j}\}_{j\neq i}, N}\\
    \mu &= \left(\frac{\partial S}{\partial N}\right)_{\{X_{j}\}, \{Y_{j}\}},
\end{align}
\end{subequations}
estos parámetros intensivos $E_{i}$, $C_{i}$ y $\mu$ pueden ser explorados para buscar formas de equilibrio y explorar la validez de los supuestos planteados acá en futuros trabajos. De momento se puede ver además de la ecuación \ref{Intensivos} que aparecen de manera natural unos procesos en los que los parámetros no cambian.

Podemos definir como un proceso isoeconómico un proceso en el cual no hubo variación en ninguno de los $dX_{i}$, de manera similar el proceso isocultural se daría cuando no hay variación en ninguno de los $dY_{i}$. Por otro lado, el proceso asociado a $dN=0$ corresponde a una dinámica sin crecimiento poblacional. Esto ocurre cuando el número de inmigrantes y nacimientos iguala al número de emigrantes y fallecimientos.

Un proceso en el cual los signos simbólicos no cambian se puede decir que es reversible y puede ser llamado isosimbólico siguiendo la nomenclatura adoptada para los casos previos. 

Otras formas de procesos pueden ser identificados cuando alguna de las variables permanece constante como se ve en la ecuación \ref{dS}, en ese sentido si la cantidad de monedas $X_{M,i}$ o la cantidad de libros publicados $Y_{L,i}$ no cambian representan unos proceso específicos. Estos procesos pueden ser identificados para cada una de las variables que caracterizan el proceso, así como para los parámetros intensivos definidos en la ecuación \ref{Intensivos}.

\section{Equilibrio Respecto a los Recursos Disponibles}
Como caso puntual para explorar la forma de equilibrio en este contexto se puede analizar una sociedad simplificada sin cambios culturales, poblacionales y como único cambio económico el de los recursos disponibles.

Sea $S$ un sistema compuesto por dos subsistemas tales que $S_{i}=S(R_{i})$, donde $R_{i}$ representa la cantidad de recursos disponibles para el sistema $i$-ésimo. Si el sistema está aislado se puede suponer que los recursos totales se conserven, de manera que:
\begin{equation}
    dR=dR_{1}+dR_{2}=0,
\end{equation}
bajo esta perspectiva se asume que los recursos no se generan, es decir que no se encuentran nuevas fuentes de ese recurso. En ese caso bajo este modelo en que estas dos sociedades solo pueden interactuar por medio del intercambio de esos recursos y bajo la suposición de aditividad de $S$ al definirla acá como un parámetro extensivo:
\begin{equation}\label{dS_2}
    dS=dS_{1}+dS_{2}=(E_{R,1}-E_{R,2})dR_{1}\geq 0,
\end{equation}
es decir que el equilibrio se alcanza cuando $dR_{1}=dR_{2}=0$ o para el estado en el cual $E_{R,1}=E_{R,2}$. El signo o la interpretación de $E_{R}=dS/dR$ señala el tipo de sociedad a modelar, por ejemplo si se establece que $E_{R}$ es positiva por definición en la ecuación \ref{dS_2} entonces la sociedad que esté tomando los recursos es la que tenga el valor de $E_{R}$ mayor. En el caso contrario si fuese negativa o fuese el inverso de alguna otra función relevante puede pensarse que la evolución se daría de manera tal que si $dR_{1}$ aumenta es porque inicialmente $E_{R,1}$ es menor.

Se puede interpretar la ecuación \ref{dS_2} como en la representación entrópica de la termodinámica. Si se establece una función $Z$ tal que $Z=Z(S,R)$ como una representación diferente puede pensarse en una forma de equilibrio de Z en relación a $R$ diferente que permita interpretar la ecuación \ref{dS_2} de otra manera, haciendo la analogía entre la representación energética y entrópica [1].

La validez de esta formulación puede explorarse al estar formulada en términos cuantitativos y pueden establecerse medidas para estimar el tamaño del sistema o su estabilidad respecto a ciertos parámetros por ejemplo se puede estudiar el crecimiento poblacional de múltiples países, calculando la fluctuación cuadrática media para establecer si pueden ser considerados como macroscópicos o no.

\section{Conclusiones}

Un modelo termodinámico de la sociedad se presenta como una primera aproximación para interpretar el comportamiento de sistemas sociales en términos de parámetros económicos y culturales medibles. Se plantea la posibilidad de establecer parámetros intensivos sin analogías a ecuaciones de estado de la termodinámica, que sirvan para estudiar posibles formas de equilibrio o de evolución hacia el equilibrio.

Algunos aspectos deberán ser explorados en futuros trabajos como la interpretación de $S$ bajo otra perspectiva y su validez como una función no decreciente para los sistemas sociales. Así como la búsqueda exhaustiva de múltiples parámetros extensivos y no extensivos que sean no decrecientes como posibles propuestas para $S$.

Los aspectos económicos y sociales fueron modelados acá por medio la diversificación estructural de bienes, recursos así como por el multimonetarismo. Otras variables económicas pueden ser exploradas para futuros trabajos y la relación de ellas con sus variables conjugadas en relación a $S$ o bajo otra formulación usando la riqueza como en [2] $W=\sum_{i=1}^{x_{m}}\lambda_{i} X_{M,i}+\sum_{i=1}^{x_{b}}p_{i}X_{B,i}$ donde los $\lambda_{i}$ representan el valor de la moneda $i$-ésima y los $p_{i}$ representan los precios de los bienes.

Finalmente las posibles aplicaciones de una teoría social como la que se presenta acá han de poder describir los resultados empíricos obtenidos por otros modelos de la sociedad. Aunque múltiples autores han abordado el problema mecánico estadístico de la construcción por colectividades aún faltan avances en los aspectos fundamentales que justifiquen la elección de ciertos enfoques.

\bibliographystyle{apalike} 
\bibliography{references} 

\end{document}